\documentclass[preprint]{aastex62}
\usepackage{amsmath}

\graphicspath{{./}{figures/}}

\received{--}
\revised{--}
\accepted{--}
\submitjournal{ApJL}

\shorttitle{A Magnetized, Moon-Forming Giant Impact}
\shortauthors{Mullen \& Gammie}

\begin{document}

\title{A Magnetized, Moon-Forming Giant Impact}

\correspondingauthor{Patrick Mullen}
\email{pmullen2@illinois.edu}

\author[0000-0003-2131-4634]{P. D. Mullen}
\affiliation{Department of Astronomy, University of Illinois at Urbana-Champaign, 1002 West Green Street, Urbana, IL, 61801, USA}

\author[0000-0001-7451-8935]{C. F. Gammie}
\affiliation{Department of Astronomy, University of Illinois at Urbana-Champaign, 1002 West Green Street, Urbana, IL, 61801, USA}
\affiliation{Department of Physics, University of Illinois at Urbana-Champaign, 1110 West Green Street, Urbana, IL 61801, USA} 



\begin{abstract}
The Moon is believed to have formed in the aftermath of a giant impact between a planetary mass body and the proto-Earth. In a typical giant impact scenario, a disk of vapor, liquid, and solid debris forms around the proto-Earth and--after possibly decades of evolution--condenses to form the Moon. Using state-of-the-art numerical simulations, we investigate the dynamical effects of magnetic fields on the Moon-forming giant impact. We show that turbulence generated by the collision itself, shear in the boundary layer between the post-impact debris field and the proto-Earth, and turbulence in the vapor component of the disk amplify the field to dynamically significant strengths. Magnetically driven turbulence promotes angular momentum transport in the protolunar disk.  Debris material is accreted onto the proto-Earth, making Moon formation less efficient, while the disk is forced to spread to larger radii, cooling at its outer edge.  Magnetic fields speed the evolution of the vapor component of the protolunar disk and hasten the formation of the Moon.

\end{abstract}



\section{Introduction} \label{sec:intro}
In the giant impact scenario for the formation of the Moon, a planetary mass impactor strikes the proto-Earth during the final stages of terrestrial planet formation \citep{hartmann+1975,cameron+1976}. After a transient impact phase that lasts $\sim$24 hr, the debris either escapes or settles into orbit about the proto-Earth forming a ``protolunar disk". The long-term evolution of the disk and the lunar assembly process are not yet well understood, but it is clear that hot vapor formed in the impact is sufficiently opaque that its radiative cooling time is $\gtrsim$10 yr.  The giant impact is dynamically complicated, involving strong shocks, melting and vaporization, self-gravity, and a nontrivial equation of state. Studies of the giant impact therefore commonly employ numerical simulations \citep[see][and references therein]{barr2016}, but simulations have yet to incorporate magnetic fields.

The notion that magnetic fields might play a role in the evolution of the protolunar disk was first proposed by \cite{charnoz+2015}, followed by \cite{carballido+2016} and \cite{gammie+2016}. Each argue that some regions of the protolunar disk could be well-coupled to any existing magnetic field. If a sufficiently strong seed field is present, the protolunar disk is unstable to the magnetorotational instability \citep[MRI,][]{balbus+1991}, which would drive turbulence, causing angular momentum transport and disk evolution.

Are seed fields hosted by the target and impactor strong enough to produce a magnetized remnant? Young, solar-mass stars have strong winds and are known to host magnetic fields far stronger than the present-day Sun \citep[e.g.,][]{johns-krull2007}, suggesting that the early Solar System was permeated by a similar, magnetized wind. All planets in the Solar System (with Venus as a potential exception) either hosted or presently host a magnetic field.  The ancient magnetic field of the Earth is not yet well-characterized: there is evidence for a field hosted by Earth as much as $\sim$4.2 Ga \citep{tarduno+2015,tarduno+2020}, although these findings are disputed \citep{weiss+2015,borlina+2020}. Remnant magnetization found in meteorites provides evidence that many, but not all, meteorite parent bodies hosted a magnetic field within a few Myr of the formation of the Solar System \citep[see][and references therein]{weiss+2017}. It is therefore plausible that the giant impact collision partners also hosted a magnetic field. It is unlikely, however, that the initial field is strong enough (several kilogauss, see below) to be dynamically important.

If the target and impactor do contain seed magnetic fields and the debris field is well-coupled, what processes might lead to field amplification? Turbulent amplification of an initial field can occur during the transient impact phase. For example, in the first hours after contact, turbulence is driven by the Kelvin-Helmholtz instability along the contact surface (i.e., at the impact site shear layer, where the colliding planets graze past one another). Turbulence amplifies the magnetic field exponentially, but amplification associated with the contact layer is difficult to numerically resolve. In a study of magnetized neutron star-neutron star mergers by \cite{kiuchi+2015}, the field strength was amplified by a factor of $\sim$10$^3$, but the total amplification was not converged at their maximum resolution.

Magnetic winding in a differentially rotating flow amplifies a magnetic field linearly in time.  The protolunar disk is centrifugally supported with Keplerian rotational frequency $\Omega_K \sim 1.2 \times 10^{-3} \left( R/R_\oplus \right)^{-3/2}  \: \mathrm{s}^{-1}$ where $R$ is disk radius (radial pressure support, however, can substantially lower the orbital frequency). Magnetic winding of radial field $B_R$ grows the toroidal field strength $B_\phi$ following $\partial_t B_\phi \sim B_R \: R \: d \Omega / dR$.  If the initial post-impact field $B_R \sim B_\phi$, then at post-impact time $t$
\begin{equation}
B_\phi \sim 60 \left(\frac{B_R}{1 \: \mathrm{G}} \right) \left( \frac{R}{2 R_\oplus}\right)^{-3/2} \left(\frac{t}{24 \: \mathrm{hr}} \right) \: \mathrm{G}.
\end{equation}
The magnetic field is dynamically important when the magnetic pressure $P_B \equiv B^2/\left(8 \pi \right)$ is comparable to the gas pressure $P_g$, i.e., $B \sim 5 \sqrt{P_g / \left( 1 \: \mathrm{bar}\right)} \: \mathrm{kG}$.  Evidently winding in the protolunar disk alone can amplify plausible initial fields to dynamically interesting strengths in a few months.  Magnetic winding is fastest where the shear rate is the largest.  In a canonical giant impact scenario, the post-impact remnant can exhibit an abrupt transition (i.e., over $\Delta R \simeq 0.5 R_\oplus$) from the rotational profile of the proto-Earth to the rotational profile of the centrifugally supported protolunar disk.  In this ``boundary layer" between the proto-Earth and protolunar disk, the shear rate can be larger than anywhere else in the post-impact remnant and hence exhibit the largest field amplifications associated with magnetic winding.  We contrast this scenario with the ``synestia" model \citep{lock2018} where the distinction between proto-Earth and protolunar disk is blurred and the transition between the rotation profile of the molten Earth and outer disk is gradual.

MRI-driven turbulence can cause an initially weak field to grow exponentially in time \citep{hgb95}.  In the absence of resistive diffusion, the protolunar disk is unstable to the MRI. The MRI is suppressed by Ohmic resistivity $\eta$ if the magnetic Reynolds number $\mathrm{Re}_\mathrm{M} \equiv H^2 \Omega / \eta \lesssim \left[B^2/ \left(4 \pi P_g \right) \right]^{-1}$, where $H$ is the disk scale height.  The Ohmic resistivity $\eta$ depends on the free electron abundance and the electron collision frequency \citep{draine2011}.  \cite{gammie+2016} anticipate that the thermal ionization of Na is the major source of free electrons in the protolunar disk and that $\mathrm{Re}_\mathrm{M} \gtrsim 10^6$ in the vapor phase, so the MRI will therefore be suppressed if $B \lesssim 3.5 \left(\mathrm{Re}_\mathrm{M} / 10^6 \right)^{-1/2} \left(P_g / \left(1 \: \mathrm{bar} \right) \right)^{1/2} \: \mathrm{G}$. As noted above, winding can  amplify an initially $\sim$1 G field above the Ohmic diffusion threshold. Then, MRI driven turbulence can amplify the field on the dynamical timescale.

To accurately model MRI driven turbulence, the most unstable mode must be resolved.   In linear theory, the MRI has maximum growth rate $\gamma_{\mathrm{max}} = 0.5 \lvert d \Omega / d \ln R \rvert$ at wavelength $\lambda_\mathrm{max} = 2 \pi \sqrt{16/15} v_A / \Omega$, where $v_A = B/ \sqrt{4 \pi \rho} \equiv \mathrm{Alfv\acute{e}n \: velocity}$ and $\rho \equiv \mathrm{mass \: density}$.  To resolve the fastest growing mode, a simulation must have linear resolution $\Delta x \lesssim \lambda_\mathrm{max} / 10$. Put differently,
\begin{equation}
B_\mathrm{midplane} \gtrsim 2 \times 10^3 \left(\frac{\Delta x}{250 \: \mathrm{km}} \right) \left(\frac{\rho_\mathrm{midplane}}{10^{-3} \: \mathrm{g/cm}^3} \right)^{1/2} \left(\frac{R}{2 R_\oplus} \right)^{-3/2} \: \mathrm{G}
\end{equation}
is the field strength required to overcome numerical diffusion in a Keplerian protolunar disk midplane. This imposes a significant, practical limit on our calculations.

Many impact scenarios for the formation of the Moon have been proposed: the ``canonical" giant impact \citep{canup+2001,canup2004}, multiple giant impacts \citep{rufu+2017,citron+2018}, ``hit-and-run" scenarios \citep{asphaug2010,reufer+2012}, impacts with a fast-spinning proto-Earth leading to the formation of the Moon inside a terrestrial ``synestia" \citep{cuk2012,lock2018}, to name a few.  We anticipate that magnetic fields could be dynamically important in all impact scenarios, as long as the post-impact remnant is (a) differentially rotating and (b) well-coupled to the field.  Differential rotation is seemingly ubiquitous in at least some regions of all post-impact structures, independent of the collision scenario.  We anticipate that high angular momentum impact scenarios, particularly the ``synestia" model, may yield post-impact structures that approach virial temperatures, facilitating near perfect coupling to any existing magnetic field.

\section{Numerical Setup}
To explore the possibility that a weak initial field is amplified by a combination of magnetic winding and turbulent growth until it is dynamically important, we have built numerical models using the grid-based magnetohydrodynamics (MHD) framework \texttt{Athena++} \citep{stone+2020}.  

In this first study, we consider a ``hit-and-run" impact scenario \citep{asphaug2010,reufer+2012} between two undifferentiated granite planets, with total mass $M_t \simeq 1.1 M_\oplus$, impactor-to-total-mass ratio $\xi \simeq 0.18$, impact angle $\theta \simeq 35^\circ$, and impact velocity $v_\mathrm{imp} \simeq 1.2 v_\mathrm{esc}$ (where $v_\mathrm{esc} \equiv \mathrm{mutual \: escape \: velocity}$). Both the target and impactor are initially in nonrotating, isentropic, hydrostatic equilibrium, with 
\begin{equation}
-\frac{1}{\rho} \frac{\partial P}{\partial r} - \frac{G M(r)} {r^2} = 0
\end{equation}
and
\begin{equation}
\frac{\partial P}{\partial r} - c_s^2 \frac{\partial \rho}{\partial r} = 0,
\end{equation}
where $r$ is spherical radius and $M(r)$ is enclosed mass. We affix a low-mass atmosphere ($M_\mathrm{atmos} \lesssim 3.6 \times 10^{-3} \: M_t$) to each of the collision partners to facilitate a transition from the planetary profiles to an ambient density and pressure, $\rho_a = 10^{-6} \: \mathrm{g/cm}^3$ and $p_a = 10^{-2} \: \mathrm{bar}$.  

Each body hosts a dipole-like magnetic field. We adopt the \cite{paschalidis+2013} current loop vector potential (in cgs units and spherical coordinates),
\begin{equation}
A_\phi = \frac{\pi r_0^2 I_0 \varpi}{c \left(r_0^2 + r^2 \right)^{3/2}} \left(1 + \frac{15 r_0^2 \left(r_0^2 + \varpi^2 \right)}{8\left(r_0^2 + r^2 \right)^2} \right),
\end{equation}
where $r_0$ is the current loop radius, $I_0$ is the current, $\varpi^2 = r^2 \sin^2 \theta$, $\theta$ is the spherical polar angle, and the subscript $\phi$ denotes the toroidal (azimuthal) component. We choose $r_0$ to be a third of the planetary body’s radius.  $A_\phi$ can be rotated to form any angle desired between the magnetic pole of the planetary body and the angular momentum vector of the collision.  All models presented in this work have a proto-Earth dipole vector aligned with, and an impactor dipole vector perpendicular to, the initial orbital angular momentum.  The magnetic field is initialized by taking the curl of the summed vector potentials of the collision partners, so that $\nabla \cdot \mathbf{B} = 0$.  $I_0$ sets the magnetic field strength. The initial dipole is characterized by the surface field strength at the magnetic poles, $B_0$.

We use an ideal MHD model with self-gravity applying a second-order accurate van Leer predictor-corrector time integrator \citep{stone+2009}, piecewise parabolic reconstruction \citep{colella+1984}, a local Lax-Friedrichs Riemann solver \citep{toro2013}, and the open-boundary condition Poisson solver of \cite{moon+2019}. The coupling of MHD to self-gravity enters through momentum and total energy source terms; these source terms are added following the momentum and total energy conserving \cite{hanawa2019} scheme, which requires two Poisson solves per timestep. Our fluid boundary conditions are outflow, with no inflow permitted. The numerical integration has Courant-Friedrichs-Lewy number $= 0.3$.  Our simulations employ the Tillotson equation of state for granite \citep{tillotson1962}.  The Tillotson equation of state models both condensed (liquid/solid) and expanded (vapor) states.  A simple mixing rule between these two regimes (i.e., a linear interpolation between condensed and expanded states) is used to describe the intermediate (liquid/vapor mixture) states.

We survey two linear resolutions ($\Delta x \simeq 175 \: \mathrm{km}$ and $\Delta x \simeq 350 \: \mathrm{km}$) and two initial field strengths ($B_0 \simeq 1 \: \mu\mathrm{G}$ and $B_0 \simeq 1 \: \mathrm{kG}$). Our mesh is Cartesian and cubic (uniform) with edge length $\sim$14 $R_\oplus$.  Linear resolutions $\Delta x \simeq 350 \: \mathrm{km}$ and  $\Delta x \simeq 175 \: \mathrm{km}$ correspond to $N_\mathrm{mesh} = 256^3$ and $N_\mathrm{mesh} = 512^3$ , respectively.  Each simulation is evolved to $\sim$240 hr post-impact.

\section{Results}

\begin{figure}[ht!]
\centering
\includegraphics[width=\textwidth]{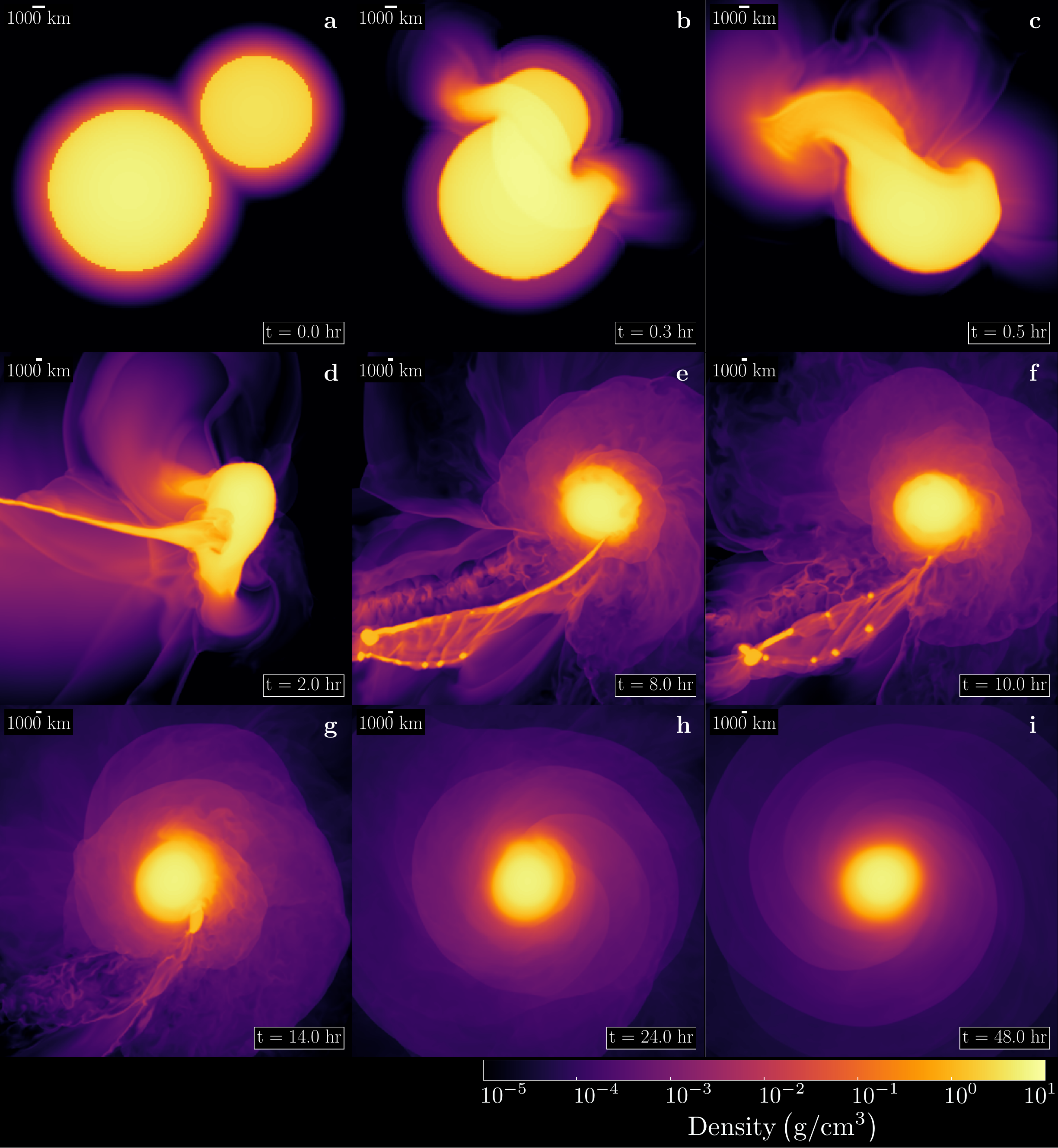}
\caption{The evolution of ($\log_{10}$) density at the collision midplane for a $B_0 \simeq 1 \: \mathrm{kG}$, $\Delta x \simeq 175 \: \mathrm{km}$ magnetized giant impact simulation.  Each panel has a unique window size, bringing attention to: (a)-(c) the shocks propagating through the planetary bodies during the collision (0-0.5 hr), (d)-(g) the tidal arm of debris (2-14 hr), and (h)-(i) the protolunar disk (24-48 hr). \label{fig:fig1}}
\end{figure}

Figure \ref{fig:fig1} follows the first $\sim$48 hr after impact for the $B_0 \simeq 1 \: \mathrm{kG}$, $\Delta x \simeq 175 \: \mathrm{km}$ model. In Figure \ref{fig:fig1}(b), two shocks form at the contact surface. One propagates through the proto-Earth and the other propagates through the impactor. The peak densities and pressures in the simulation lie along the shock propagating through the proto-Earth. In between the shocks, the specific internal energy (a proxy for temperature) is high ($e \simeq 2.3 \times 10^{11} \: \mathrm{erg/g} = 2.3 \times 10^{7} \: \mathrm{J/kg}$). Shear at the contact surface leads to amplification of the magnetic field. The layer is unstable to the Kelvin-Helmholtz instability, but for rolls to grow within the brief lifetime of the shear layer, $\Delta x \simeq 50 \: \mathrm{km}$ is needed \citep{mullen+2020}. As the initial encounter finishes (Figure \ref{fig:fig1}(c)), the proto-Earth is in a transient, distorted state. In Figures \ref{fig:fig1}(d)-\ref{fig:fig1}(g), from $\sim$2 to $\sim$14 hr post-impact, impactor- and target-derived material form a tidal arm. The impact launches a wave that wraps around the surface of the proto-Earth. Field amplification continues as magnetic field lines are stretched and wound by the wave and tidal arm. Bound debris falls back to the proto-Earth and wraps around it, adding mass to the nascent protolunar disk. About 24 hr after initial contact, the debris has settled into a disk. The wave of material launched by the impact continues to propagate around the surface of the proto-Earth with a $\sim$4 hr period. This wave drives spiral shocks into the protolunar disk \citep[c.f.,][]{wada+2006,canup+2013}.

By 48 hr post-impact, the magnetic field is strongest in the boundary layer, where the shear rate is largest.  The interior of the proto-Earth preserves The boundary layer is stable to the MRI because the MRI requires an outwardly decreasing angular velocity profile. Earlier work has shown, however, that the boundary layer is unstable to acoustic waves that steepen into outwardly propagating spiral shocks \citep{belyaev+2012a,belyaev+2012b,belyaev+2013}.  At 48 hr post-impact, the boundary layer magnetic field at the midplane is $\sim$8 kG (comparable to the field strength in a medical MRI machine), with plasma $\beta \equiv P_g/P_B \simeq 600$. In the disk midplane, field strength declines with increasing radius, reaching $\sim$350 G at $R \sim 4 R_\oplus$. The pressure decreases outward so that midplane $\beta$ is roughly constant ($\sim$10$^4$) at $R\sim2-6 R_\oplus$.

\begin{figure}[ht!]
\centering
\includegraphics[width=\textwidth]{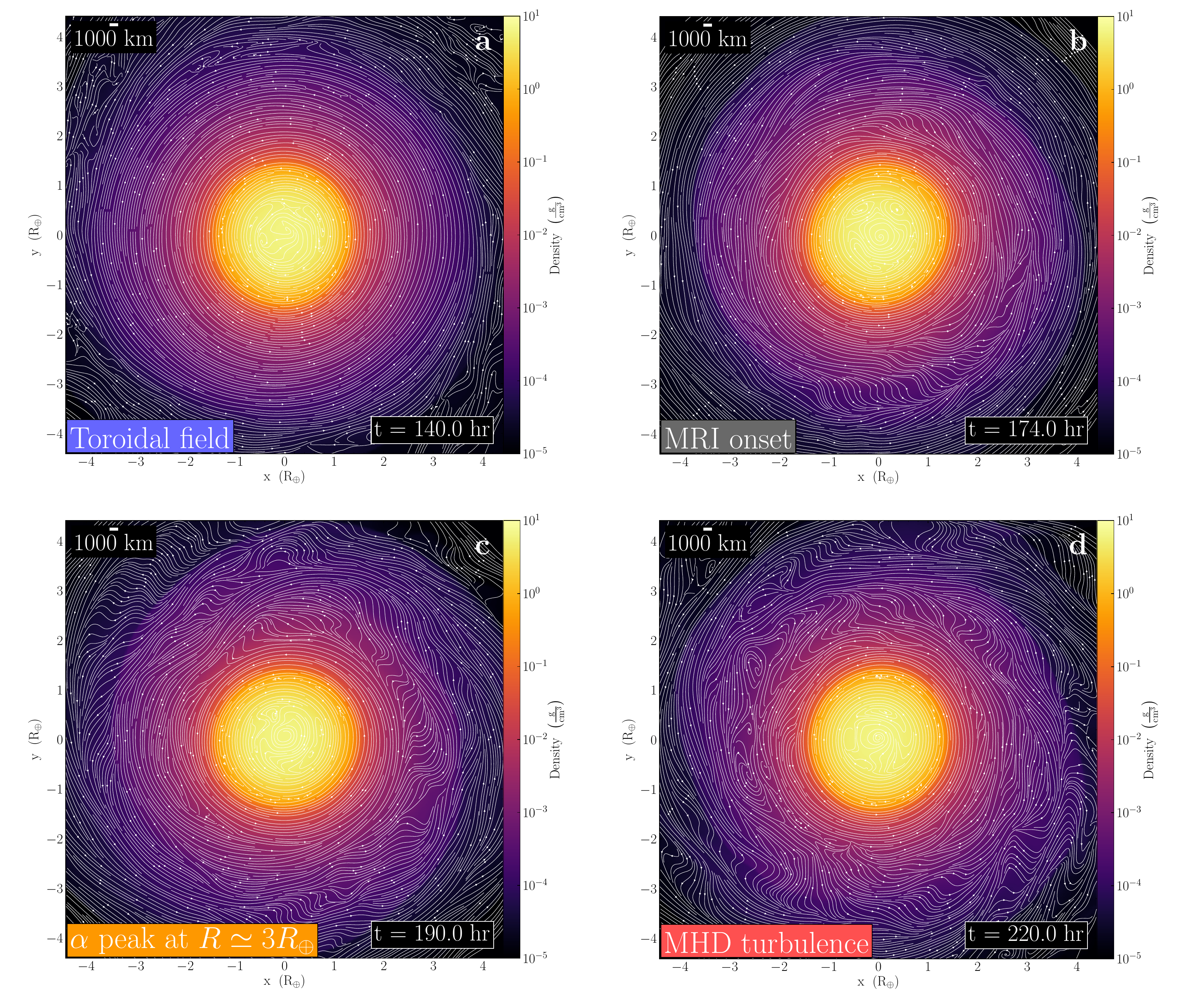}
\caption{Logarithmic density distribution of the protolunar disk midplane with overplotted lines corresponding to the $x-$ and $y-$components of the magnetic field for a $B_0 \simeq 1 \: \mathrm{kG}$, $\Delta x \simeq 175 \: \mathrm{km}$ magnetized giant impact simulation. (a) Quasi-equilibrium protolunar disk, $\sim$140 hr post-impact, hosting a mainly toroidal magnetic field. (b) The onset of the MRI, $\sim$174 hr post-impact. (c) The peak of the dimensionless $\alpha$ parameter around $R \simeq 3 R_\oplus$, $\sim$190 hr post-impact. (d) Full MHD turbulence in the protolunar disk, $\sim$220 hr post-impact. \label{fig:fig2}}
\end{figure}

Beyond 120 hr post-impact, turbulence fueled by the collision itself has mostly dissipated, leaving behind a magnetized, differentially rotating protolunar disk.  Figure \ref{fig:fig2}(a) shows the magnetic field configuration (overplotted lines) at 140 hr post-impact, superposed on the density field. We find that the natural outcome of a magnetized Moon-forming giant impact is a mainly toroidal (azimuthal) magnetic field.

By 174 hr post-impact, the MRI begins to develop in the disk (Figure \ref{fig:fig2}(b)).  Magnetic field lines begin to depart from a toroidal configuration. In Figure \ref{fig:fig2}(c), nontoroidal components of the field are comparable to the toroidal components; the field is disordered. By 220 hr post-impact, only the boundary layer retains a toroidal field (Figure \ref{fig:fig2}(d)). The remainder of the protolunar disk has been engulfed by MRI-driven turbulence.

\begin{figure}[ht!]
\centering
\includegraphics[width=\textwidth]{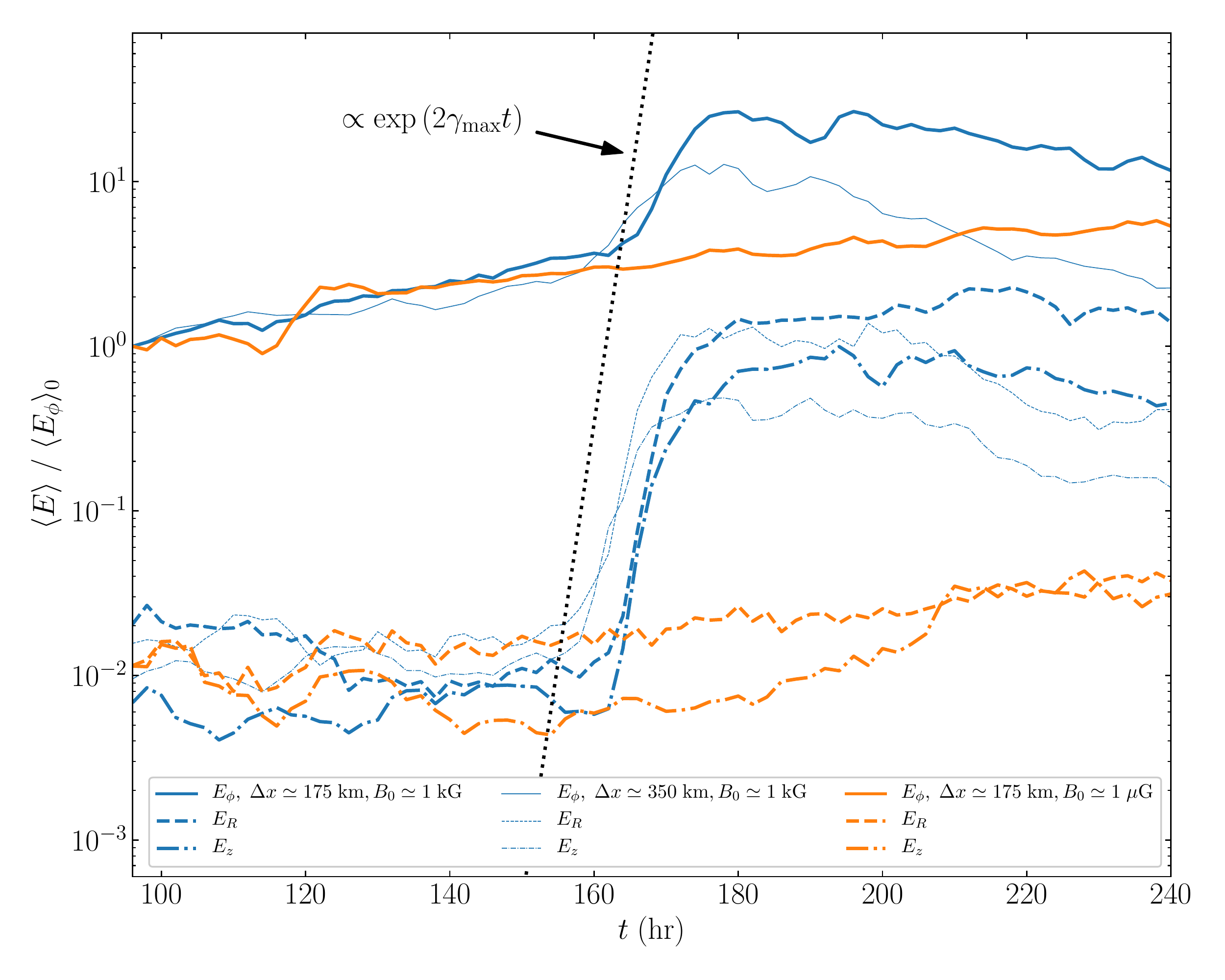}
\caption{Toroidal ($E_\phi$, solid),  radial ($E_R$, dashed),  and vertical ($E_z$, dot-dashed) magnetic energies (volume averaged in cylindrical annuli described in the main text) for the magnetized calculations at two initial field strengths, $B_0 \simeq 1 \: \mathrm{kG}$ (blue) and $B_0 \simeq 1 \: \mu\mathrm{G}$ (orange). For the kilogauss run, we present two linear resolutions, $\Delta x \simeq 175 \: \mathrm{km}$ (bold) and $\Delta x \simeq 350 \: \mathrm{km}$ (fine). We over plot the growth rate expected for the most unstable MRI wavelength, $\gamma_\mathrm{max}$ (black, dotted). \label{fig:fig3}}
\end{figure}

We expect the components of the magnetic energy (toroidal $E_\phi \equiv B_\phi^2 / 8\pi$, radial $E_R \equiv B_R^2 / 8\pi$, and vertical $E_z \equiv B_z^2 / 8\pi$) in the protolunar disk to grow exponentially upon MRI onset in the $B_0 \simeq 1 \: \mathrm{kG}$ calculations, with growth rate $\gamma \sim 2 \gamma_\mathrm{max}$ (magnetic energy is quadratic in $B$).  The $B_0 \simeq 1 \: \mu\mathrm{G}$ calculations should not exhibit an observable exponential phase, as the resolved MRI wavelengths are unstable with growth rate $\gamma \sim \gamma_\mathrm{max} \left(\lambda_\mathrm{max}/\lambda \right)$, i.e., they grow too slowly. For all initial field strengths, we expect a steady increase in toroidal magnetic energy due to magnetic winding, with field strengths growing linearly in time. Figure \ref{fig:fig3} shows the magnetic energy associated with the three components of the field for the kilogauss and microgauss (kinematic) runs, averaged over cylindrical annuli whose origins are at the center of the proto-Earth with inner radii $R_\mathrm{inner}$, outer radii $R_\mathrm{outer}$, and heights $h$. Selecting annuli with $\left(R_\mathrm{inner},R_\mathrm{outer},h \right)_{\Delta x \simeq 175 \: \mathrm{km}}$ = $\left(3 R_\oplus,4 R_\oplus,0.5 R_\oplus \right)$ and $\left(R_\mathrm{inner},R_\mathrm{outer},h \right)_{\Delta x \simeq 350 \: \mathrm{km}}$ = $\left(3.85 R_\oplus,4.85 R_\oplus,0.5 R_\oplus \right)$ guarantees that the sampling regions are outside the boundary layer, focused at the midplane, and have equivalent $\gamma_\mathrm{max}$ for both model linear resolutions.  For the $B_0 \simeq 1 \: \mathrm{kG}$ runs, we observe exponential growth in all three components of the magnetic energy at MRI onset $\sim$170 h post-impact, with energy growth rate $\gamma \sim 2 \gamma_\mathrm{max}$.  As anticipated, the $B_0 \simeq 1 \: \mu\mathrm{G}$ run exhibits no detectable exponential growth: the radial and vertical components of the magnetic energy remain roughly constant beyond $\sim$100 hr post-impact and the toroidal component grows steadily due to magnetic winding.

\begin{figure}[ht!]
\centering
\includegraphics[width=\textwidth]{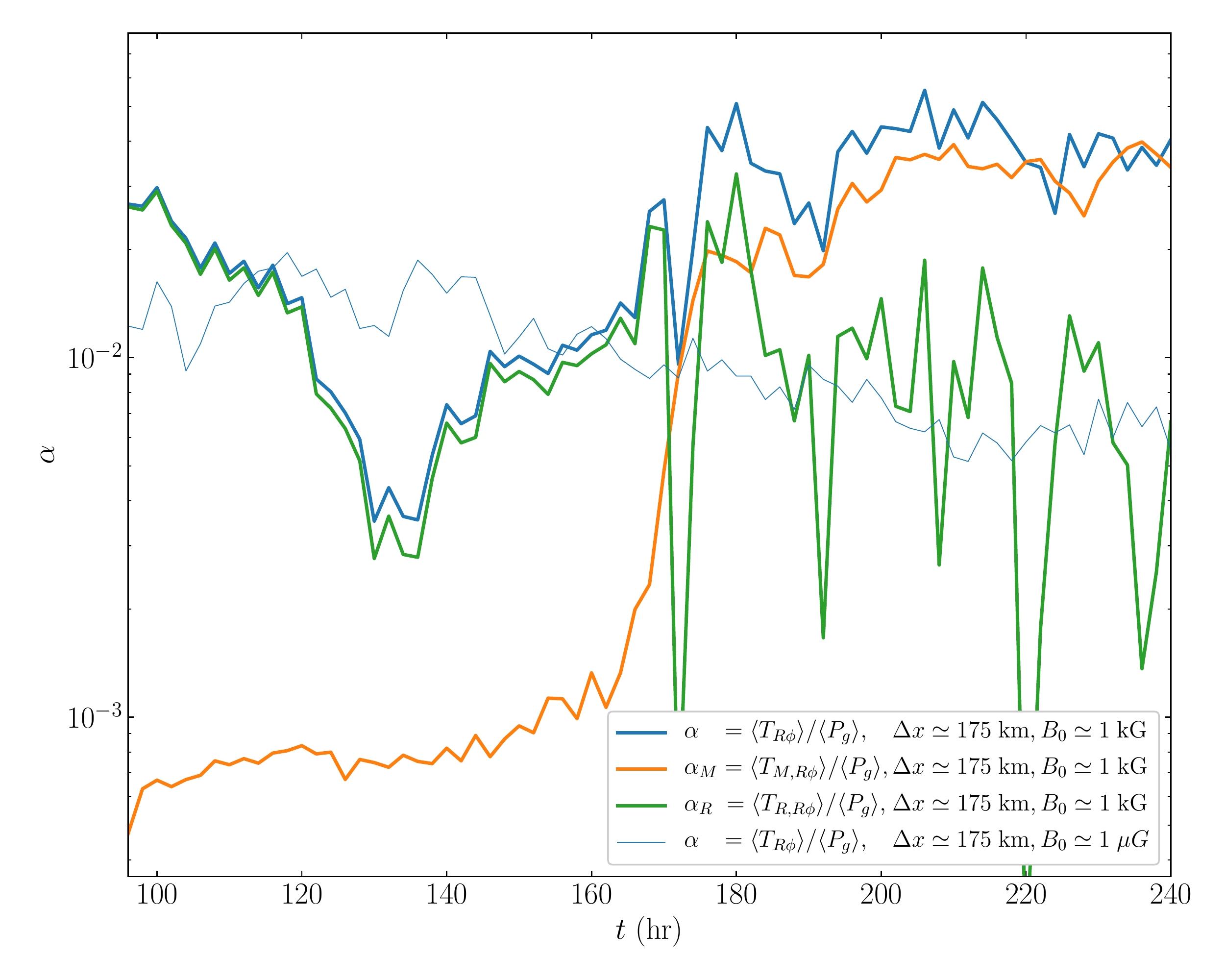}
\caption{The dimensionless $\alpha$ parameter (volume averaged in cylindrical annuli described in the main text) for the magnetized calculations at two initial field strengths, $B_0 \simeq 1 \: \mathrm{kG}$ (blue, bold) and $B_0 \simeq 1 \: \mu\mathrm{G}$ (blue, fine).  For the kilogauss run, we also show contributions to $\alpha$ from Maxwell stresses $\alpha_M = \langle T_{M,R \phi} \rangle / \langle P_g \rangle$ (orange) and Reynolds stresses $\alpha_R = \langle T_{R,R \phi} \rangle / \langle P_g \rangle$ (green). \label{fig:fig4}}
\end{figure}

Disk turbulence leads to angular momentum transport. The angular momentum flux density is proportional to the sum of the Reynolds and Maxwell stress. The volume-averaged Reynolds stress is
\begin{equation}
\left< T_{R,R\phi} \right> = \left< \rho v_R \delta v_\phi \right>
\end{equation}
where $v_R$ is the radial velocity, $\delta v_\phi = v_\phi - \overline{v_\phi}$, and $\overline{v_\phi}$ is the azimuthally averaged orbital velocity. The volume-averaged Maxwell stress is
\begin{equation}
\left< T_{M,R\phi} \right> = -\left< \frac{B_R B_\phi}{4 \pi} \right>
\end{equation}
The ratio of the volume-averaged total stress to the volume-averaged gas pressure gives a dimensionless measure of the strength of turbulent transport \citep{shakura+1973},
\begin{equation}
\alpha = \frac{\left<T_{R\phi} \right>}{\left<P_g \right>} = \frac{\left< \rho v_R \delta v_\phi - B_R B_\phi/\left(4 \pi \right) \right>}{\left<P_g \right>}
\end{equation}
Figure \ref{fig:fig4} presents the evolution of $\alpha$ for the kilogauss and microgauss (kinematic) runs in the same cylindrical annuli described above.  Prior to MRI onset (t$\sim$140 hr) in the $B_0 \simeq 1 \: \mathrm{kG}$, $\Delta x \simeq 175 \: \mathrm{km}$ model, the ratio of Maxwell to Reynolds stresses is $\sim$1/3, with $\alpha \sim 3 \times 10^{-3}$. After MRI onset, the Maxwell stress overtakes the Reynolds stress and increases by more than an order of magnitude. After peaking ($\sim$190 hr, Figure \ref{fig:fig2}(c)), the stresses equilibrate at $\alpha \sim 5 \times 10^{-2}$.  We note that Maxwell stresses do not contribute to $\alpha$ in the $B_0 \simeq 1 \: \mu\mathrm{G}$, $\Delta x \simeq 175 \: \mathrm{km}$ model. The microgauss run exhibits a steadily declining $\alpha$ as turbulence generated from the impact is dissipated.  At $\sim$240 hr post-impact, $\alpha$ in the kilogauss run is nearly an order of magnitude larger than $\alpha$ in the microgauss run.

\begin{figure}[ht!]
\centering
\includegraphics[width=\textwidth]{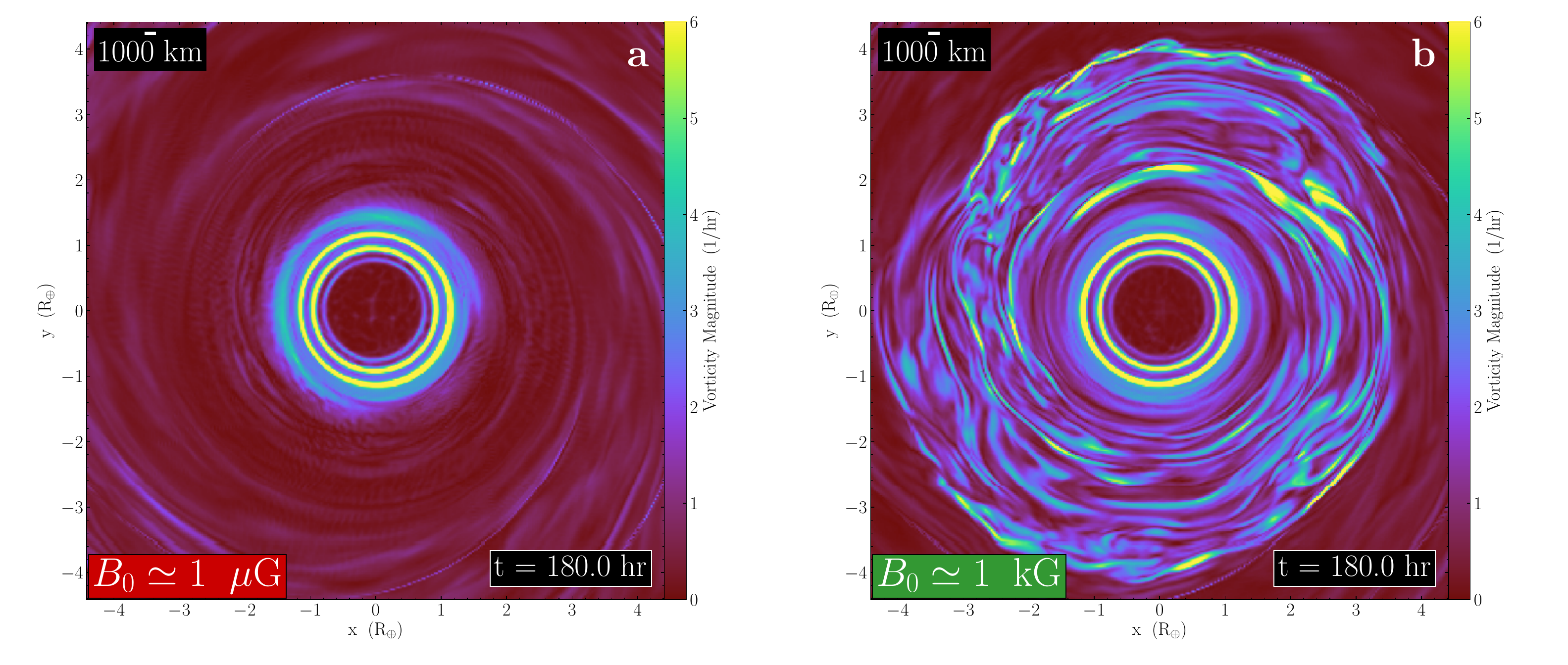}
\caption{Vorticity magnitude in the protolunar disk midplane at $\sim$180 hr post-impact for models with linear resolution $\Delta x \simeq 175 \: \mathrm{km}$ and initial field strengths (a) $B_0 \simeq 1 \: \mu\mathrm{G}$ and (b) $B_0 \simeq 1 \: \mathrm{kG}$. Evidently the weak field model is nearly laminar, while the strong field model is turbulent. \label{fig:fig5}}
\end{figure}

Turbulence is expected to lead to growth of vorticity in the protolunar disk.  Figure \ref{fig:fig5} presents the magnitude of the vorticity at the disk midplane for the microgauss (Figure \ref{fig:fig5}(a)) and the kilogauss (Figure \ref{fig:fig5}(b)) $\Delta x \simeq 175 \: \mathrm{km}$ runs. The snapshot is taken $\sim$180 hr post-impact, after onset of the MRI.  In the kinematic run, the vorticity is non-negligible only in the boundary layer. In the kilogauss run, the MRI-active disk is turbulent, with vorticity exceeding the local orbital angular frequency by as much as an order of magnitude.

\section{Discussion and Conclusions}

What are the implications of a magnetized giant impact for the formation of the Moon? We can confirm that the disk becomes fully turbulent, which supports the scenario proposed by \cite{gammie+2016}.  It is likely that the vapor component of the protolunar disk evolves rapidly, on a timescale
\begin{equation}
t_\mathrm{spread} \simeq 60 \left(\frac{\alpha}{10^{-2}} \right)^{-1} \left(\frac{R}{2 R_\oplus} \right)^{3/2} \left(\frac{H}{R} \right)^{-2} \: \mathrm{hr}.
\end{equation}
A large fraction of a magnetized protolunar disk will likely accrete onto the proto-Earth, making Moon formation less efficient than in an unmagnetized scenario. The remainder of the disk will quickly spread to larger radius and cool adiabatically. Eventually, adiabatic cooling at the outer edge of the disk will lead to reduced free electron abundance, increasing resistivity and magnetic decoupling.  

This scenario differs markedly from that of \cite{thompson+1988} in that the introduction of magnetic fields reduces the disk spreading timescale from $\sim$100 yr to $\sim$100 hr.  Disk material that evades accretion onto the proto-Earth and spreads to large radii \citep[decoupling at $R \sim 10 R_\oplus$,][]{gammie+2016} will supply the reservoir of material forming the Moon, challenging the generally accepted picture that Moon-formation occurs just outside the Roche radius $\simeq 2.9 R_\oplus$.

Our models leave a number of open questions.  First, we have not addressed how magnetic fields influence mixing between impactor- and proto-Earth derived material, aside from our result that a sufficiently strong seed field yields vigorous magnetic turbulence in the protolunar disk.  We anticipate that mixing will proceed on a timescale $t_\mathrm{mix} \simeq t_\mathrm{spread}$.  Second, we do not know the effects of including an iron core in the collision partners.  \cite{hosono+2016} find that the inclusion of iron cores can affect the ejecta masses/velocities following the impact and the final structure of the protolunar disk.  Third, we do not yet know how the disk will evolve with a more sophisticated, tabulated equation of state \citep[e.g.,][]{melosh2007}. The \cite{tillotson1962} equation of state is known to have drawbacks, particularly in modelling material vaporization \citep[see][]{stewart+2019}.  Nevertheless, our results should be unbiased to material thermodynamics as magnetic winding and the magnetorotational instability (MRI) are incompressible phenomena. Fourth, our models do not consider the role of finite conductivity in the proto-Earth and protolunar disk; by integrating the equations of ideal MHD, we assume that the flow is everywhere perfectly coupled to the magnetic field.  This assumption may be violated inside the proto-Earth, close to the surface of the protolunar disk, or at large radii where temperatures are low.  However, cooling is inefficient in the protolunar disk.  Assuming a radiative protolunar disk, \cite{gammie+2016} find a cooling time $t_\mathrm{cool} >$ 10 yr, partially attributed to the large opacity of silicate vapor.  The heating timescale of the protolunar disk is $t_\mathrm{heat} \simeq \left[\alpha \Omega_K \right]^{-1}$.  Seemingly, cooling can be entirely neglected and the disk will undergo runaway heating until the bulk of the protolunar disk is well-coupled to the field, where the assumption of ideal MHD is then nearly valid.  In this scenario, disk spreading is the only viable mechanism of cooling the protolunar disk and promoting decoupling from the field.

Our study makes it plausible that the proto-Earth and protolunar disk are strongly magnetized in the first few weeks following the Moon-forming giant impact. There may already be geochemical evidence for a magnetically active protolunar disk: \cite{nie+2019} identify a stark contrast in $^{87}\mathrm{Rb}/^{85}\mathrm{Rb}$ between Earth and lunar samples. They conclude that the depletion of such moderately volatile elements stems from the accretion of a vaporous layer in the protolunar disk onto the proto-Earth. The turbulent viscosity needed for their scenario could be realized through a magnetically coupled disk.

\acknowledgments

We are indebted to the \texttt{yt} project \citep{turk+2011} which made the visualizations (and much of the analysis) in this work possible.  We thank Yufeng Du, Jonah Miller, Ben Ryan, the \texttt{Athena++} collaboration (particularly Matt Coleman, Josh Dolence, Kyle Felker, Sanghyuk Moon, and Jim Stone) for their help and comments.  We additionally thank the referee, S\'ebastien Charnoz, for a thoughtful report that greatly benefited the manuscript. This work is supported by the National Aeronautics and Space Administration under Grant Award 80NSSC19K0515 issued through the Emerging Worlds Program. We gratefully acknowledge supercomputer time on NASA’s Pleiades (allocation HEC-SMD-18-1885), TACC’s stampede2 at the University of Texas at Austin (allocation TG-AST170024), and NCSA’s Blue Waters at the University of Illinois at Urbana-Champaign (allocation ILL\_bawj).

%

\vspace{5mm}
\facilities{Pleiades, stampede2, Blue Waters}

\software{\texttt{Athena++} \citep{stone+2020},
          \texttt{yt} \citep{turk+2011},  
          }

\bibliographystyle{aasjournal}
\bibliography{ads}



\pagebreak

\pagebreak

\pagebreak

\pagebreak

\end{document}